\documentclass[aip,apl,12pt]{revtex4-2}
\usepackage{graphicx}
\usepackage{color}

\begin{document}

%\preprint{}

\title{Gate-controlled Supercurrent in Ballistic InSb Nanoflag Josephson Junctions}



\author{Sedighe Salimian}
%\thanks{}
\affiliation{NEST, Istituto Nanoscienze-CNR and Scuola Normale Superiore, Piazza San Silvestro 12, 56127 Pisa, Italy}

\author{Matteo Carrega}
%\thanks{}
\affiliation{CNR-SPIN, Via Dodecaneso 33, 16146, Genova, Italy}

\author{Isha Verma}
%\thanks{}
\affiliation{NEST, Istituto Nanoscienze-CNR and Scuola Normale Superiore, Piazza San Silvestro 12, 56127 Pisa, Italy}

\author{Valentina Zannier}
%\thanks{}
\affiliation{NEST, Istituto Nanoscienze-CNR and Scuola Normale Superiore, Piazza San Silvestro 12, 56127 Pisa, Italy}

\author{Michał P. Nowak}
%\thanks{}
\affiliation{AGH University of Science and Technology, Academic Centre for Materials and Nanotechnology, Krakow, Poland}

\author{Fabio Beltram}
%\thanks{}
\affiliation{NEST, Istituto Nanoscienze-CNR and Scuola Normale Superiore, Piazza San Silvestro 12, 56127 Pisa, Italy}

\author{Lucia Sorba}
%\thanks{}
\affiliation{NEST, Istituto Nanoscienze-CNR and Scuola Normale Superiore, Piazza San Silvestro 12, 56127 Pisa, Italy}

\author{Stefan Heun}
\email[]{stefan.heun@nano.cnr.it}
%\thanks{}
\affiliation{NEST, Istituto Nanoscienze-CNR and Scuola Normale Superiore, Piazza San Silvestro 12, 56127 Pisa, Italy}


\date{\today}

\begin{abstract}
High-quality III-V narrow band gap semiconductor materials with strong spin-orbit coupling and large Land\'e g-factor provide a promising platform for next-generation applications in the field of high-speed electronics, spintronics, and quantum computing. Indium Antimonide (InSb) offers a narrow band gap, high carrier mobility, and a small effective mass, and thus is very appealing in this context. In fact, this material has attracted tremendous attention in recent years for the implementation of topological superconducting states supporting Majorana zero modes. However, high-quality heteroepitaxial two-dimensional (2D) InSb layers are very difficult to realize owing to the large lattice mismatch with all commonly available semiconductor substrates. An alternative pathway is the growth of free-standing single-crystalline 2D InSb nanostructures, the so-called nanoflags. Here we demonstrate fabrication of ballistic Josephson-junction devices based on InSb nanoflags with Ti/Nb contacts that show gate-tunable proximity-induced supercurrent up to 50~nA at 250~mK and a sizable excess current. The devices show clear signatures of subharmonic gap structures, indicating phase-coherent transport in the junction and a high transparency of the interfaces. This places InSb nanoflags in the spotlight as a versatile and convenient 2D platform for advanced quantum technologies.
\end{abstract}

%\keywords{}

\maketitle


%\section{Introduction}

Today a great interest revolves around the possibility to create and manipulate new states of matter with topological properties. This stems mostly from the intrinsic robustness of topological states against local perturbation and the ensuing relevance for quantum computing architectures.\cite{Haldane2017,Nayak2008} Hybrid superconductor/semiconductor heterostructures represent a promising platform in which topological properties can emerge.\cite{Prada2020,Stern2013,Fornieri2019,Moehle2021}

In this context, Indium Antimonide (InSb) has attracted much attention recently. InSb has a narrow band gap ($\sim 0.23$ eV).\cite{Chen2021a,Moehle2021,Qu2016} It also has a very high bulk electron mobility ($7.7 \times 10^4$~cm$^2$/(Vs))\cite{Mata2016,Ashley1995} and a small effective mass ($m^* = 0.018 \: m_e$),\cite{Sladek1957,Qu2016,Mata2016,Ke2019,Vries2019,Lei2021} which are both important requirements for high-speed and low-power electronic devices.\cite{Ashley1995,Zutic2004} Finally, it also exhibits a strong spin-orbit interaction and a large Land\'e g-factor ($\left| g^* \right| \sim 50$,\cite{Mata2016,Lei2021}) and thus it is useful for spintronics applications \cite{Qu2016,Zutic2004} and for the creation of hybrid structures hosting topological states, like Majorana zero-modes. Indeed, the first signatures compatible with Majorana bound states were reported in InSb nanowires coupled to a superconductor,\cite{Prada2020,Mourik2012} which has triggered strong efforts to improve the quality of hybrid systems based on InSb nanowires.\cite{Nilsson2012,Li2016,Gazibegovic2017,Guel2017,Guel2018,Gill2018,Sestoft2018,Pendharkar2021}

Besides one-dimensional nanowires, two-dimensional (2D) InSb structures also attract great attention, owing to their inherent design flexibility.\cite{Moehle2021,Qu2016,Lei2021} Indeed, InSb 2D electron gases \cite{Ke2019} and the related ternary compound InSbAs \cite{Moehle2021} have recently been proposed as a platform for topological superconductivity,\cite{Lesser2021} and ballistic superconductivity was demonstrated in InSb quantum wells.\cite{Ke2019} However, the growth of high-quality heteroepitaxial 2D InSb layers is still a challenge owing to their large lattice mismatch with common semiconductor substrates. Besides, such quantum wells are reported to suffer from instabilities due to the Si dopants.\cite{Lei2021} A possible strategy to circumvent these problems consists in growing free-standing 2D InSb nanostructures on nanowire stems, because nanowires yield efficient relaxation of elastic strain along the nanowire sidewalls, when lattice-mismatched semiconductor systems are integrated. To emphasize their free-standing 2D shape, such nanostructures are often referred to as nanosails, nanosheets, nanoflakes, or nanoflags. However, until today only a few studies were reported on the growth and the electrical transport properties of such InSb nanoflags (NF).\cite{Mata2016,Pan2016,Gazibegovic2017,Vries2019,Zhi2019,Zhi2019a,Verma2020,Verma2021,Chen2021a}

InSb NFs were first reported in 2016 by M.~de la Mata \textit{et al.}\cite{Mata2016} their growth being based on molecular beam epitaxy (MBE). There, the authors attributed the 2D geometry of the NFs to a single twinning event in the otherwise pure zinc blende structure of the InSb sample, and four-terminal electrical measurements revealed an electron mobility greater than 12000~cm$^2$/(Vs). D. Pan \textit{et al.}~used Ag-assisted MBE to grow free-standing 2D single-crystalline InSb NFs.\cite{Pan2016} Hall-bar devices were then fabricated that showed ambipolar behavior and electron mobility of 18000~cm$^2$/(Vs). The same group also demonstrated functional InSb NF Josephson junctions (JJ) with Al and Nb contacts.\cite{Zhi2019} Furthermore, in the devices with Nb contacts authors reported evidence of the coexistence of the quantum Hall effect and proximity-induced superconductivity in the InSb NFs.\cite{Zhi2019a} Quantum transport and quantum-dot geometries in such nanostructures were also demonstrated very recently.\cite{Kang2019,Xue2019,Chen2021,Chen2021a} S.~Gazibegovich \textit{et al.}~combined selective-area growth with the vapour-liquid-solid mechanism in metal organic vapour phase epitaxy, leading to the formation of InSb NFs thanks to the development of a twin-plane boundary.\cite{Gazibegovic2017,Gazibegovic2019} The same group reported evidence for crossed Andreev reflections in JJ made from such flakes.\cite{Vries2019}

Also our group demonstrated the growth of single-crystal, free-standing InSb NFs, initially on InAs nanowires, using a directional growth technique based on chemical beam epitaxy (CBE).\cite{Verma2020} There, NF size was limited by the flexibility of the InAs nanowires, which led to a bending of the stem and the resulting loss of the orientation for the asymmetric 2D growth. In order to overcome this issue, we recently optimized the growth of InSb NFs.\cite{Verma2021} In particular, the InSb NFs were grown on sturdy tapered InP nanowires, which did not bend and allowed to grow larger NF with the same directional-growth approach. This strategy allowed to obtain InSb NFs of $\left( 2.8 \pm 0.2 \right)$~$\mu$m length, $\left(470 \pm 80\right)$~nm width, and $\left(105 \pm 20\right)$~nm thickness.\cite{Verma2021}

The resulting NFs are large enough to fabricate Hall bars with length-to-width ratios enabling precise electrical characterization. An electron mobility of $29500$~cm$^2$/(V s) was measured at a carrier concentration $n = 8.5 \times 10^{11}$~cm$^{-2}$ at $4.2$~K.\cite{Verma2021} The electron elastic mean free path $\ell_e$ reached values of $500$~nm, which favorably compares with present literature.\cite{Mata2016,Gazibegovic2019,Vries2019}

Here, we report on the fabrication and characterization of JJ devices based on these InSb NFs and provide evidence of ballistic superconductivity. We employ Ti/Nb contacts in InSb JJ devices and show gate-tunable proximity-induced supercurrent at 250~mK and a sizable excess current. The devices also show clear signatures of subharmonic gap structures, indicating phase-coherent transport in the junction and highly transparent interfaces. Our results indicate InSb NF as a promising platform for the study of topological superconductivity.

%\section{Results}

\begin{figure}[t]
	\includegraphics[width=0.5\textwidth]{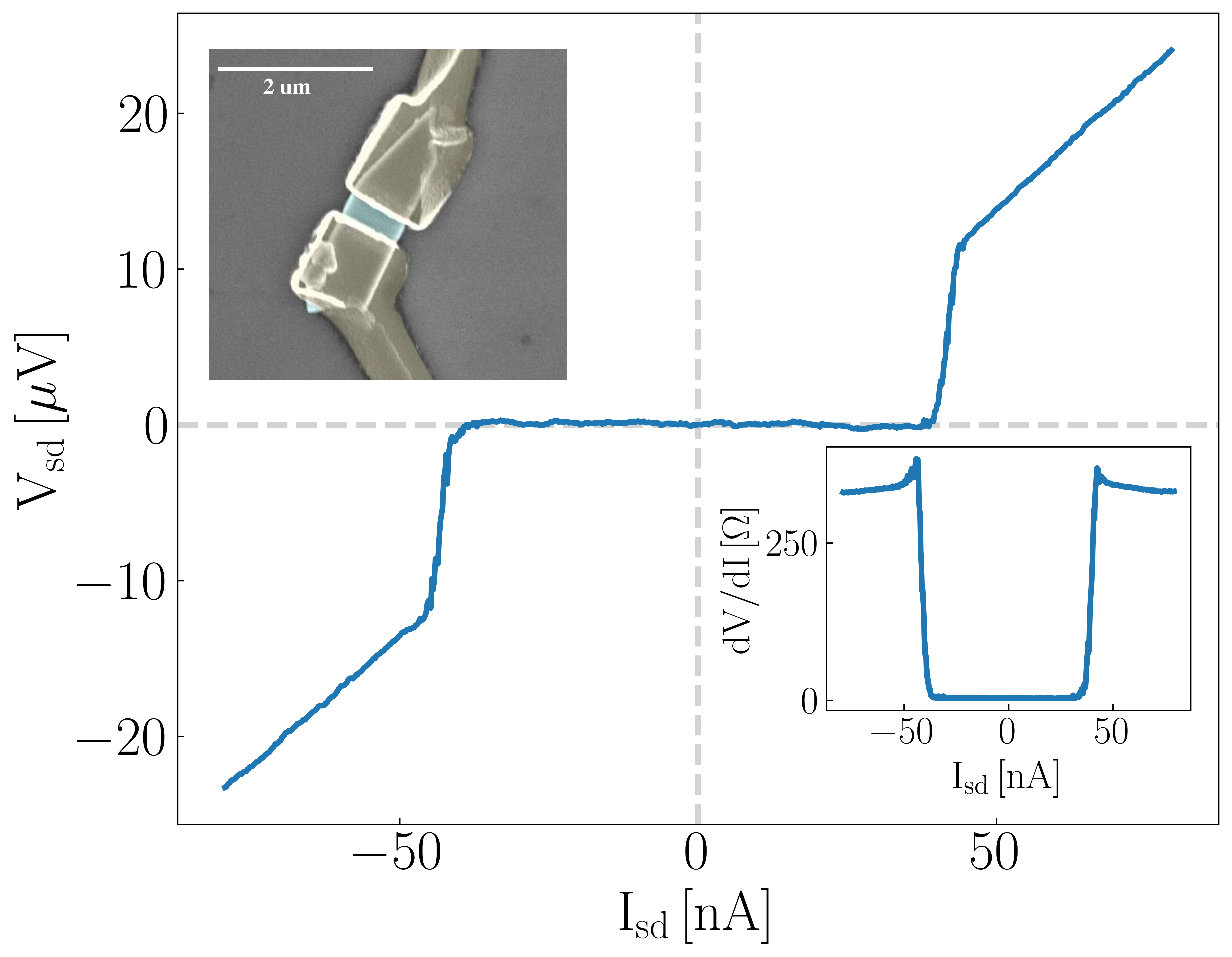}
	\caption{\label{IV} DC voltage drop $V_{sd}$ as a function of bias current $I_{sd}$. A supercurrent of $\sim 50$~nA is observed. The lower right inset shows the differential resistance $dV/dI$ measured simultaneously by lock-in technique. $V_{bg} = 30$~V,  $T = 250$ mK. $B = 6$~mT applied to compensate for the residual magnetization of the cryostat. The upper left inset shows an SEM image of the device structure. Scale bar 2 $\mu$m.}
\end{figure}

The upper left inset to Fig.~\ref{IV} shows a SEM image of the device investigated in this work (for consistent results from other devices, see the Supplementary Material). In brief, a 100~nm-thick InSb NF was transferred mechanically on a SiO$_2$/Si substrate and contacted with 10/150~nm Ti/Nb (more details on device fabrication are provided in the Supplementary Material). The interelectrode spacing between the two superconductors, i.e., the length of the normal (N) region, is $L = 200$~nm, while its width is $W = 700$~nm. Standard transport characterization yields a mean-free path of $\ell_e \sim 500$~nm \cite{Verma2021} for the N region, greater than the junction length $L$. These numbers place the device in the ballistic regime.

The critical temperature of the superconducting leads was determined to be $T_c = 8.44$~K, from which the bulk gap can be computed using the theory by Bardeen, Cooper, and Schrieffer (BCS):\cite{Grosso2000} $\Delta = 1.76 k_B T_c = 1.28$~meV, consistent with values of Nb superconducting contacts previously reported in literature.\cite{Rohlfing2009,Amado2013,Gunel2014,Zhi2019a,Guiducci2019,Guiducci2019a,Carrega2019} The induced superconducting coherence length \cite{Lee2018,Vries2019,Zhi2019,Banszerus2021} is $\xi_s = \hbar v_F / \Delta$, with $v_F$ the Fermi velocity in the N region ($v_F = 1.5 \times 10^6$~m/s) and $\Delta$ the gap in the superconductor. Here, $\xi_s \sim 750$~nm $> L$, so the device is in the short junction regime. Equivalently, the Thouless energy \cite{Freericks2005,Lee2018} $E_{Th} = \hbar v_F / L = 4.9$~meV $> \Delta$.

Figure~\ref{IV} shows a typical voltage - current ($V - I$) characteristics obtained at $T = 250$~mK and $V_{bg} = 30$~V. The device displays well-developed dissipationless transport thus demonstrating proximity-induced superconductivity in the InSb NF. As the bias current exceeds the critical value of $\sim 50$~nA, a sudden jump in the measured voltage to a dissipative quasiparticle branch is observed, indicating that the JJ switches from the superconducting to the normal state, with a resistance of $\sim 330$~$\Omega$. Current sweeps in opposite directions show negligible hysteresis, i.e., switching and retrapping current are the same, so that in the following, we shall use switching current and critical current as synonyms. Consistently, the switching current is larger than the intrinsic thermal current noise $\delta I_{th}$ of the junction \cite{Likharev1979,Zhi2019} $\delta I_{th} = 2 e k_B T / \hbar$; here $\delta I_{th} = 10.5$~nA. The lower right inset to Fig.~\ref{IV} shows the differential resistance $dV/dI$ measured using a lock-in amplifier together with the $V-I$ curve. Data clearly show that the differential resistance is zero in the supercurrent branch of the device. Zhi et al.~report a supercurrent of 20 nA at 10 mK in Nb/InSb nanoflag SNS junctions.\cite{Zhi2019} We attribute the improved numbers reported here mainly to a higher mobility of the nanoflags and progress in device fabrication.

\begin{figure}[t]
	\includegraphics[width=0.5\textwidth]{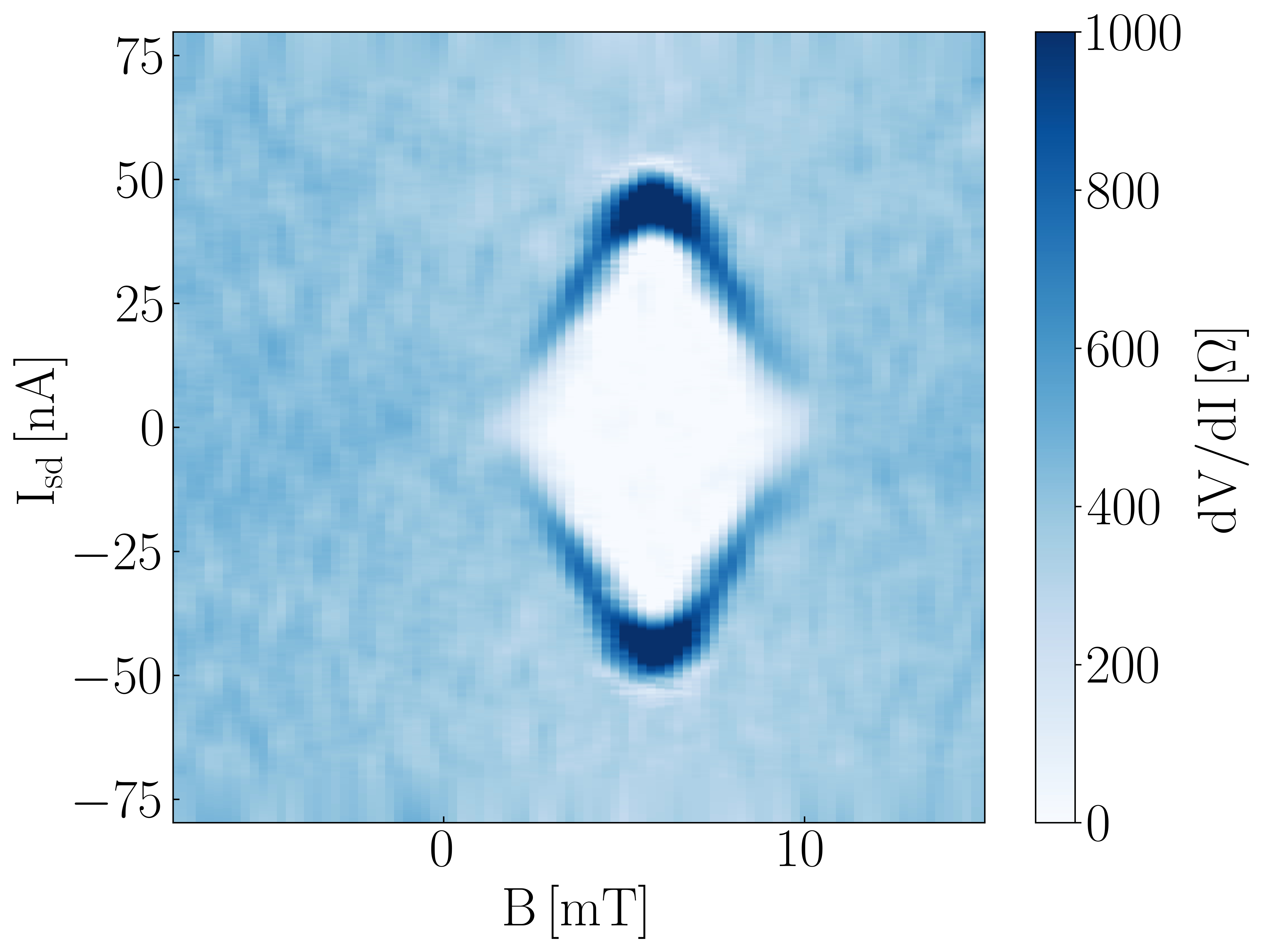}
	\caption{\label{FH} 2D colormap of differential resistance $dV/dI$, obtained by numerical derivation of measured DC $V - I$ curves, plotted vs.~current bias $I_{sd}$ and magnetic field $B$, measured at $V_{bg} = 40$~V and $T = 250$ mK.}
\end{figure}

Superconducting quantum interference was observed in the dependence of the supercurrent on a magnetic field applied perpendicularly to the sample plane (Fig.~\ref{FH}). Supercurrent maximum is obtained for $B_0 = 6$~mT instead of the expected maximum at zero B field. This small offset can be attributed to a residual magnetization in the cryostat. Applying higher or lower magnetic fields, the suppercurrent symmetrically decreases, until for $\left| B - B_0 \right| > 5.2$~mT it is suppressed. The modulation of the critical current by quantum interference is one of the hallmarks of the Josephson effect. The shape of the curve resembles a Fraunhofer pattern with only the central lobe, i.e., without side-lobes. We have verified that in a magnetic field range of $B \le 50$~mT, no side-lobes appear. De Vries et al.~have studied similar InSb NF JJ and report an even-odd Fraunhofer pattern.\cite{Vries2019} Thus, the intensity of the first side-lobes might be anomalously small which precludes their observation in our experiment. On the other hand, such anomalous magnetic interference patterns, with a monotonous decay, were reported previously in similar geometries \cite{Heida1998,Rohlfing2009,Amado2013,Kim2016} and were attributed to geometric factors.\cite{Heida1998,Barzykin1999,Sheehy2003,Cuevas2006,Cuevas2007,Bergeret2008,Carillo2008,Rohlfing2009} Besides, the magnetic flux through the junction is $\Phi = B \cdot A$, with A the junction area, $A = W (L + 2 \lambda_L)$.\cite{Chiodi2012}  Here $\lambda_L$ is the London penetration depth of Nb. Thus, the smaller the channel width $W$, the higher the value of $B$ required to reach $\Phi = \Phi_0$. According to Rohlfing et al.,\cite{Rohlfing2009} larger values of $B$ more strongly suppress $I_c(B)$ via a dramatic reduction of the amplitude of Andreev reflections. Future measurements on devices with larger W/L ratio might help to clarify whether the missing lobes in Fig.~\ref{FH} are due to geometric factors.

\begin{figure}[t]
	\includegraphics[width=\textwidth]{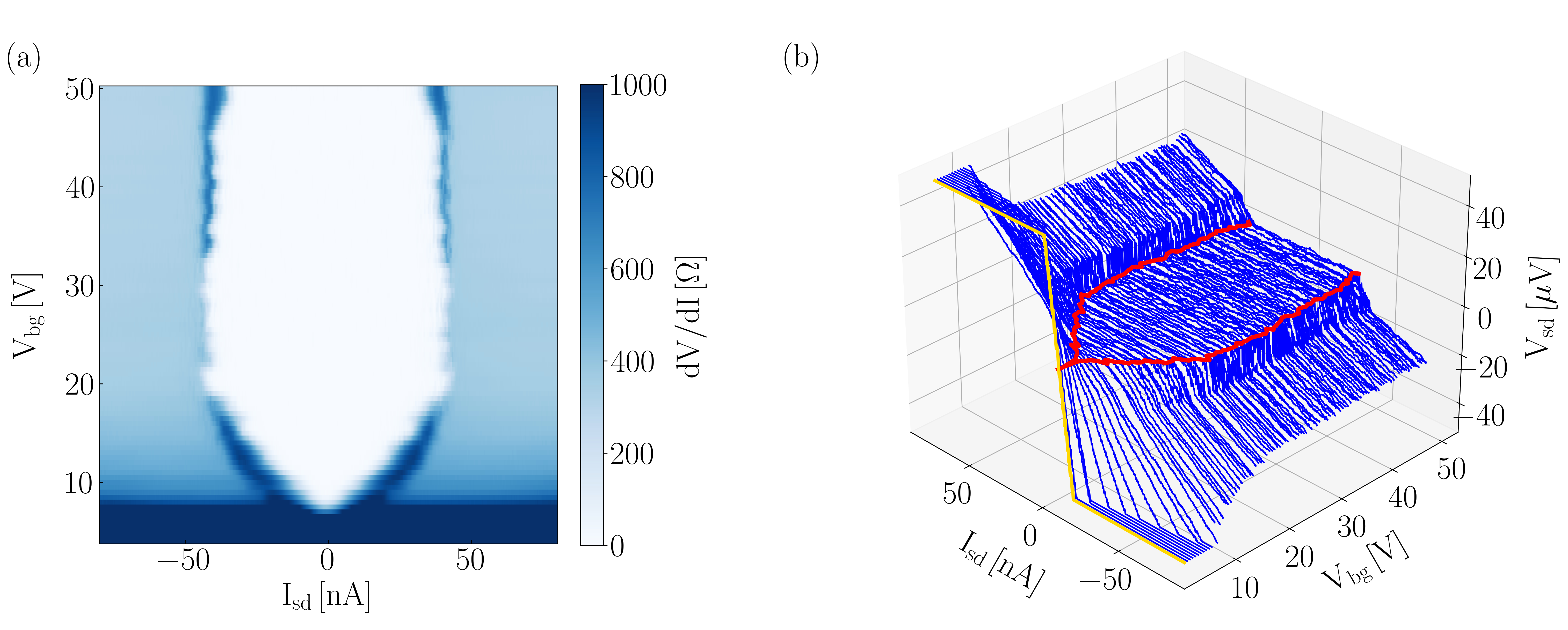}
	\caption{\label{BG} (a) Color-scale plot of differential resistance $dV/dI$, measured by lock-in technique, as a function of current bias $I_{sd}$ and back gate voltage $V_{bg}$. (b) 3D plot shows the trend of simultaneously measured DC $V-I$ curves at different back gate voltages. The first curve (at lowest back gate voltage) is highlighted in yellow. Red lines indicate the transition between the superconductive and the dissipative regime. The supercurrent increases with increasing back gate voltage above pinch off. $T = 250$ mK. $B = 6$~mT applied to compensate for the residual magnetization of the cryostat.}
\end{figure}

InSb NFs are n-type semiconductors, and the carrier concentration in the NF can be tuned by an applied gate voltage.\cite{Verma2021} Figure~\ref{BG}(a) shows that global gate modulation can also be employed to control the magnitude of the supercurrent maximum. The figure shows the differential resistance $dV/dI$ of the device as a function of current bias $I_{sd}$ and back gate voltage $V_{bg}$. The central white region represents the zero-resistance supercurrent branch. In a range of gate voltages from $20$ to $50$~V, the supercurrent is approximately constant at $\sim 50$~nA. Decreasing gate voltage below $20$~V, the supercurrent decreases, until it disappears for $\sim 5$~V. This demonstrates the gate voltage control of the supercurrent magnitude and that the device implements a Josephson Field Effect Transistor (JoFET).\cite{Clark1980,Chrestin1994,Akazaki1996,Bezuglyi2017,Wen2019} The corresponding $V-I$ curves as a function of back gate voltage are shown in Fig.~\ref{BG}(b) as line plots. While the $V-I$ curve for $V_{bg} = 10$~V still shows a zero slope at the origin, already for $V_{bg} = 7$~V the $V-I$ curve is essentially linear with an Ohmic behavior. The gate voltage dependence of the switching current is shown in Fig.~\ref{BG}(b) as red line and confirms that below $V_{bg} = 20$~V the switching current decreases. The normal resistance (the slope of the $V-I$ curves in the normal branch) displays an opposite behavior: below $V_{bg} = 20$~V it increases significantly, from $\sim 330$~$\Omega$ to above $10$~k$\Omega$. The product of switching current and normal resistance, $I_{sw} \cdot R_n$, is approximately constant at 15 $\mu$V in a wide range from $V_{bg} = 10$~V to $50$~V (see Fig. S1 in the Supplementary Material), while it drops to zero when the switching current becomes zero. Similar results were reported by Zhi et al.\cite{Zhi2019}
\begin{figure}[t]
	\includegraphics[width=0.5\textwidth]{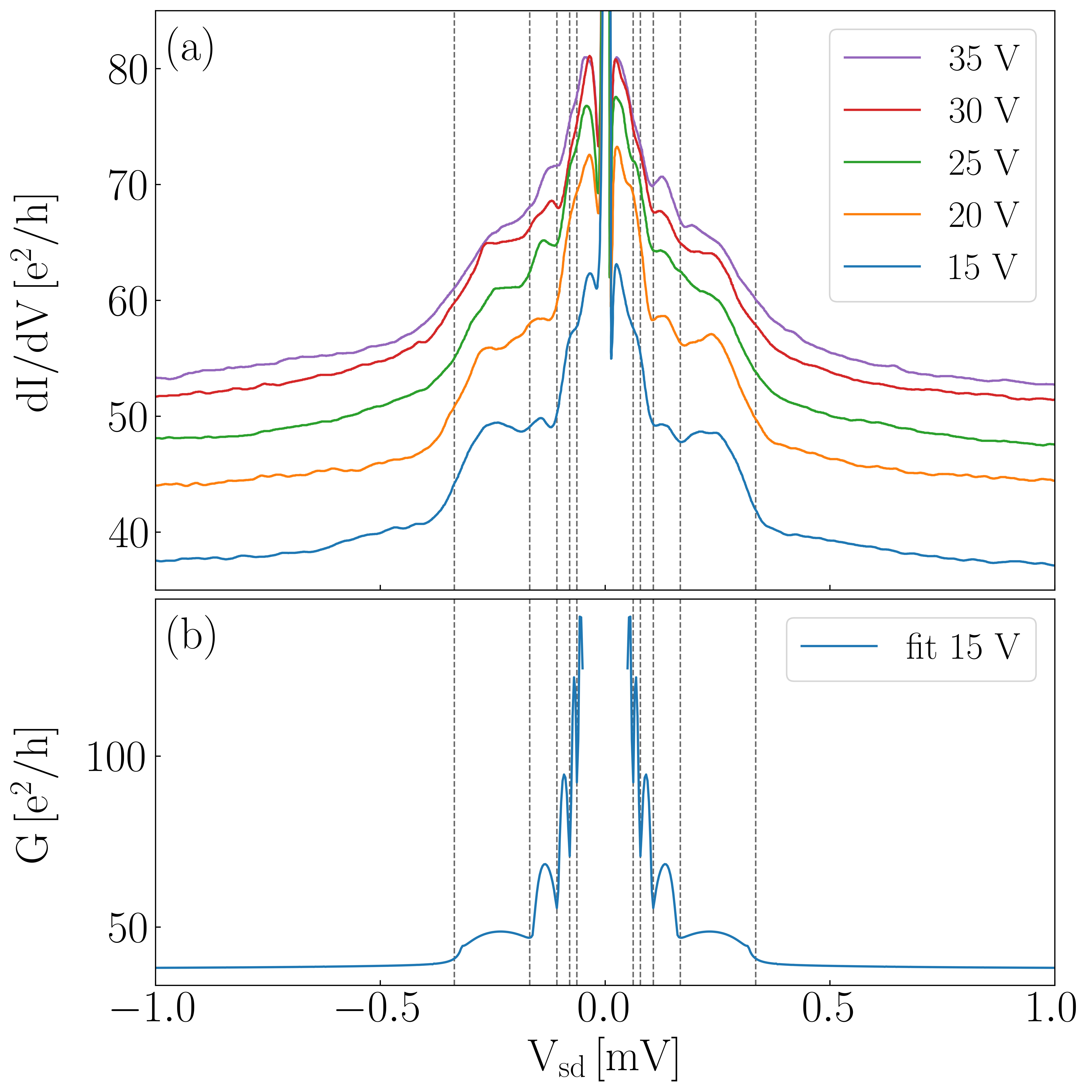}
	\caption{\label{MAR} (a) Differential conductance $dI/dV$, measured by lock-in technique, as a function of source-drain voltage $V_{sd}$ for several back gate voltages $V_{bg}$, as indicated in the legend. The dashed lines indicate minima in differential conductance caused by multiple Andreev reflections ($n = \pm 1$ to $n = \pm 5$). $T = 250$ mK. $B = 6$~mT applied to compensate for the residual magnetization of the cryostat. (b) Conductance line trace obtained from a coherent scattering model versus source-drain voltage $V_{sd}$. The theoretical fit of the data for $V_{bg} = 15$ V yields a transmission $Tr = 0.94$ and a value of the induced gap of $\Delta^* = 160$~$\mu$eV.}
\end{figure}

Next, we characterize the dissipative regime. Figure~\ref{MAR}(a) shows subharmonic gap structures in the differential conductance that can be attributed to multiple Andreev reflections (MARs). The peak present at $V_{sd} = 0$~V corresponds to the superconductive state. On the other hand, above $V_{sd} \sim \pm 0.8$ mV, the differential conductance becomes constant and is equal to the inverse of the normal resistance, $R_n^{-1}$. Between these two extrema, the differential conductance $dI/dV$ displays characteristic singularities (minima and maxima), which represent the subharmonic gap structures.\cite{Blonder1982,Klapwijk1982,Octavio1983,Flensberg1988} Their presence is a signature of the high transparency of the interfaces between S and N regions. The positions of these MAR singularities follow the equation $e V_n = 2 \Delta^* / n$, with $n = 1,2,3,...$ and $\Delta^*$ the induced gap in the N region. Most commonly, the position of the maxima in the differential conductance has been analyzed,\cite{Rohlfing2009,Deng2012,Li2016,Gharavi2017,Carrega2019,Ke2019,Vries2019,Zhi2019,Zhi2019a,Moehle2021} but recently it was pointed out that for highly transparent junctions, the MAR resonances appear as minima in the differential conductance.\cite{Kjaergaard2017}

In order to estimate the junction transparency and the induced gap, we used a simple scattering model that assumes fully-coherent transport across a multimode JJ (see Ref.~\cite{Bardas1997} and the Supplementary Material) and that has been applied to reproduce MAR traces of nanowire junctions.\cite{Heedt2021} Thus, the experimental curves were compared to optimized theoretical MAR conductance traces. One example is shown in Fig.~\ref{MAR}(b). The best agreement between experiment and theory is obtained for a junction model with 40 modes of transparency $Tr = 0.94$ and an induced gap of $\Delta^* = 160$~$\mu$eV \footnote{The next mode in the fit, the 41st mode, has $Tr = 0.15$ and thus does not contribute to the transport.}. The dashed vertical lines in Fig.~\ref{MAR} highlight the series from $n = \pm 1$ to $n = \pm 5$, showing an excellent agreement with the predicted behavior for the minima in differential conductance in a highly transparent junction.\cite{Kjaergaard2017} Results of the analysis for the other back gate voltages are very similar. Consistently, while the transparency of the modes and the induced gap do not change with back gate voltage, the number of modes does: from less than 30 at $V_{bg} = 10$~V to more than 50 at $V_{bg} = 30$~V. The obtained value of the induced gap, $\Delta^* = 160$~$\mu$eV, is smaller than the value extracted from the measurement of the critical temperature. We note that similar values for the induced gap $\Delta^*$ have been reported for InSb nanowires proximitized by Nb~\cite{Deng2012} and by NbTiN.\cite{Mourik2012}

\begin{figure}[t]
	\includegraphics[width=0.5\textwidth]{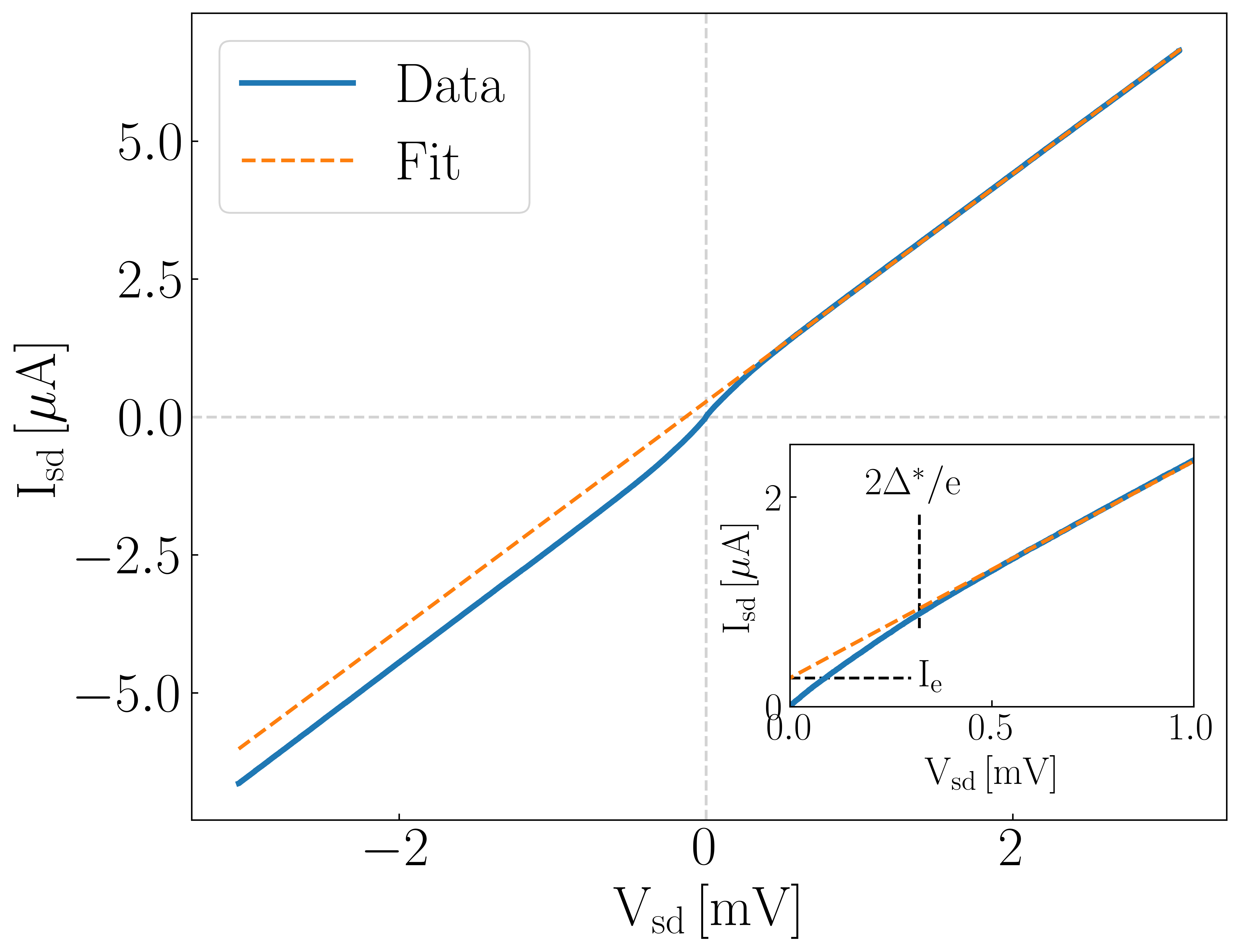}
	\caption{\label{Exc} $I-V$ characteristics measured in DC. The linear fit (dashed line) of the $I-V$ curve for $V_{sd} > 2 \Delta^* / e$ gives the excess current $I_e$ as the intercept at zero bias voltage. The inset displays the excess current in a smaller bias voltage range. $T = 250$ mK and $V_{bg} = 40$~V. $B = 6$~mT applied to compensate for the residual magnetization of the cryostat.}
\end{figure}

Besides differential conductance data, we recorded also DC traces ($I-V$ curves). Figure~\ref{Exc} shows representative example obtained with $V_{bg} = 40$~V. From a linear fit of the part of the curve at high bias, with $\left| V_{sd} \right| \ge 2 \Delta^*$, where Andreev reflections are completely suppressed, we obtain an excess current of $I_e = 265 \pm 12$~nA and a normal resistance of $R_n = 481 \pm 3$~$\Omega$. We add that the results do not change significantly if we consider only the voltage range $\left| V_{sd} \right| \ge 2$~meV. These numbers result in a product $I_e \cdot R_n = 127 \pm 7$~$\mu$V, an important figure of merit for weak links.\cite{Marsh1994} From $V_{bg} = 40$~V to $20$~V, the excess current is almost constant, while it decreases for smaller back gate values, to $\sim 50$~nA at $V_{bg} = 5$~V. The normal resistance displays the opposite behavior: below $V_{bg} = 20$~V it increases significantly to about $3.1$~k$\Omega$ at $V_{bg} = 5$~V. As shown in the Supplementary Material, the product of excess current and normal resistance, $I_e \cdot R_n$, remains approximately constant at $137 \pm 19$~$\mu$V over the whole range of back gate voltages explored, i.e., from $V_{bg} = 5$~V to $40$~V. 

%\section{Discussion}

We should like to analyse the superconducting-gap values as obtained from the critical temperature of the superconductor and the observed multiple Andreev-reflection features. The measured critical temperature $T_c = 8.44$~K is close to the reported value for bulk Nb and the resulting value of the gap $\Delta = 1.28$~meV is in good agreement with values reported for JJs with Nb contacts.\cite{Rohlfing2009,Amado2013,Gunel2014,Gharavi2017,Zhi2019a, Guiducci2019, Guiducci2019a,Carrega2019} Thus, we attribute the observed critical temperature to a switching of the Nb film from the superconducting to the normal state.

On the other hand, several groups reported gap values extracted from an analysis of MAR features that were smaller than the BCS-like gap of the superconducting leads.\cite{Deng2012,Ke2019,Pendharkar2021} Kjaergaard \textit{et al.} investigated multiple Andreev reflections in an InAs quantum-well heterostructure with epitaxial Al.\cite{Kjaergaard2017} From an analysis of the MAR features, authors obtained a gap value smaller than the BCS-like gap $\Delta$ of Al, and showed that this is due to an induced gap $\Delta^* < \Delta$ in the quantum well covered by Al. Andreev reflections of particles in the uncovered region occur at the (vertical) effective interface to the covered region with gap $\Delta^*$ in the quantum well. Since these reflections occur within the InSb crystal, the transparency of the process is high (here  $Tr = 0.94$).

Another relevant effect is the proximity-effect transfer from the Nb into InSb via the thin Ti film. Drachmann \textit{et al.}~studied the proximity-effect transfer from a NbTi film into an InAs quantum well via a thin epitaxial Al layer.\cite{Drachmann2017} They found that the induced gap $\Delta^*$ was increased by the NbTi film compared to samples with just the Al film, but it was still smaller than the BCS-like gap of NbTi. This implies that the proximity effect transfer can be weakened by an intermediate superconducting layer with smaller $T_c$. We recall that the reported $T_c$ value for bulk Ti is $\sim 0.5$~K,\cite{Webb2015,Rumble2020} i.e., much smaller than the $T_c$ of Nb. The combination of both effects is likely able to explain the observed value of the induced gap in our experiments.

The transparency of the (horizontal) interfaces between the superconducting electrodes and the InSb can be estimated using the theory of Aminov et al.,\cite{Aminov1996} which measures the transparency via the dimensionless parameter $\gamma_B$ ($\gamma_B = 0$ for perfectly transparent interfaces, larger $\gamma_B$ for larger barriers). Using the BCS gap of Nb, we obtain $\gamma_B = 12.5$. Considering that the presence of the Ti film will slightly reduce the gap by an (unknown) amount, $\gamma_B \approx 10$. This indicates a small transparency of the interface, consistent with the fact that the induced gap is much smaller than the BCS gap of Nb. On the other hand, Kjaergaard et al.\cite{Kjaergaard2017} and Baumgartner et al.\cite{Baumgartner2021} report $\gamma_B \sim 1$, consistent with their use of epitaxial Al/InAs heterostructures, which are known to have highly transparent interfaces.

Theory predicts for JJ at $T = 0$ that the product $I_{c} \cdot R_n$ is a constant proportional to the gap, $I_{c} \cdot R_n = \alpha \Delta^* / e$, with the prefactor $\alpha$ a constant of order unity.\cite{Tinkham,Ambegaokar1963,Kulik1975,Likharev1979,Beenakker1991,Mayer2019} Here $I_{c} \cdot R_n = 15$ $\mu$V is only about 10\% of $\Delta^* / e = 160$ $\mu$V. Such a reduction is frequently observed in experiment \cite{Doh2005,Li2016,Zhi2019} and has been attributed to a premature switching of the junction due to thermal activation.\cite{Tinkham,Xiang2006} On the other hand, excess current is due to Andreev reflections and thus depends primarily on the transparency of the (vertical) interface between the covered and uncovered parts of the semiconductor,\cite{Mayer2019} which is high. Consequently, a large product $I_eR_n \approx \Delta^*/e$ is observed, close to the theoretical value of 8/3 $\Delta^*/e$ for ballistic junctions.\cite{Flensberg1988,Mayer2019}
%\section{Conclusions}

In summary, we have fabricated JJ devices with InSb NFs as normal region and Ti/Nb as superconducting contacts. The high electron mobility and large mean free path of the InSb NFs yielded ballistic transport across the normal region of the junction. We showed Josephson coupling between superconductor and semiconductor, as demonstrated by the zero-resistance supercurrent of $\sim 50$~nA and the observation of MARs. Analysis of the MAR traces indicates a very high transparency of the interfaces. We also observe a sizable excess current. Our results show that free-standing 2D InSb NF on InP stems, thanks to their defect-free zinc blende crystal structure,\cite{Verma2021} are a suitable material platform for fabrication of quantum devices. Considering also their strong spin-orbit interaction and their large Land\'e g-factor, we envision the use of these structures in future studies towards topological superconductivity.

\section*{Supplementary Material}

See the Supplementary Material for Extended Methods and Additional Data.

\section*{Author's Contributions}

I.~Verma, V.~Zannier, and L.~Sorba grew the InSb NF. S.~Salimian fabricated the devices. S.~Salimian and S.~Heun carried out the experiments. M.~P.~Nowak carried out the numerical simulations. S.~Salimian, M.~Carrega, and S.~Heun analyzed the data and wrote the manuscript, with strong input from all co-authors.

\begin{acknowledgments}

We gratefully acknowledge helpful discussions with Herv\'e Courtois. We thank Daniele Ercolani for his help with the growth of the NF. This research activity was partially supported by the SUPERTOP project, QUANTERA ERA-NET Cofound in Quantum Technologies (H2020 grant No. 731473), and by the FET-OPEN project AndQC (H2020 grant No. 828948). MPN acknowledges support by the National Science Centre (NCN), agreement number UMO-2020/38/E/ST3/00418.

\end{acknowledgments}

\section*{Data Availability Statement}

The data that support the findings of this study are available from the corresponding author upon reasonable request.

\section*{Conflicts of Interest}

The authors have no conflicts to disclose.

\bibliography{JJ}

\end{document}